\documentclass[%
 reprint,
nofootinbib,
 amsmath,amssymb,
 aps,
prc
]{revtex4-2}

\usepackage{graphicx}
\usepackage{dcolumn}
\usepackage{bm}
\usepackage{hyperref}
\usepackage[mathlines]{lineno}
\usepackage{xcolor}

\begin{document}


\title{Convergence properties of $T'$-Expansion Scheme:\\ Hadron Resonance Gas and Cluster Expansion Model}

\author{Micheal Kahangirwe}
 \email{mkahangi@central.uh.edu}
 \author{Claudia Ratti}%
 \email{cratti@central.uh.edu}
\author{Volodymyr Vovchenko}%
 \email{vvovchen@central.uh.edu}
\affiliation{%
 Department of Physics, University of Houston, Houston, TX 77204, USA
}%
\author{Irene Gonzalez} 
\author{Jorge A. Muñoz}
\affiliation{
Department of Physics, University of Texas at El Paso, TX 79968, USA 
}%


\date{\today}

\begin{abstract}

In this study, we assess the effectiveness and robustness of the recently proposed $T'$-expansion scheme for expanding the equation of state of strongly interacting matter to finite density, by comparing its performance relative to the conventional Taylor expansion method in various effective QCD models.
We use baryon number density and its susceptibilities to calculate the expansion coefficients in the $T'$-expansion scheme with and without the Stefan-Boltzmann limit correction.
Our methodology involves comparing truncation orders to exact solutions to assess the scheme’s accuracy.
We utilize Ideal, Excluded Volume, and van der Waals formulations of the Hadron Resonance Gas (HRG) model at low temperatures, and the Cluster Expansion Model at higher temperatures.
Our findings indicate that the $T'$-expansion scheme offers superior convergence properties near and above the chiral crossover temperature, where the chiral-criticality-inspired scaling $(\partial/ \partial T)_{\mu_B} \sim (\partial^2/\partial \mu_B^2)_T$ holds. 
However, it shows limited improvement in the HRG models, indicating that it may not be the most suitable choice for describing the hadronic phase. 
\end{abstract}
\maketitle

\section{\label{Introduction }INTRODUCTION\protect\\ }

The attempts to describe strongly interacting matter and map out its phase diagram have progressed over the past decade. It is well established that hadrons melt into a soup of quarks and gluons at high temperature and density, a finding that was announced by experiments at RHIC \cite{STAR:2005gfr,PHENIX:2004vcz,PHOBOS:2004zne,BRAHMS:2004adc} almost two decades ago. First principle calculations from lattice QCD \cite{Aoki:2006we,Cheng:2006qk,Bhattacharya:2014ara} suggest a smooth crossover transition between hadronic and partonic matter at vanishing baryon density. At extremely high density, the smooth crossover is expected to evolve into a line of first-order phase transition with a critical point. The search for the critical point is at the core of the second Beam Energy Scan (BES II) at RHIC, and future facilities such as FAIR and JPARC~\cite{CBM:2016kpk,Durante:2019hzd,Fukushima:2020yzx}. On the theory side, effective methods and extrapolation techniques, such as Taylor expansion \cite{Gavai:2003mf,Guenther:2017hnx,HotQCD:2018pds,Borsanyi:2022soo,Bollweg:2022fqq,Bollweg:2022rps} and analytical continuation from
imaginary chemical potential \cite{Bellwied:2015rza,Borsanyi:2020fev} 
are employed to push the description to finite density, for which first principle Monte Carlo simulations are hindered by the sign problem.  
The standard Taylor expansion suffers from limitations due to the restricted availability of higher-order lattice susceptibilities, unphysical oscillations in thermodynamic observables, and finite radius of convergence, which render it unreliable at high density where the critical point is predicted. 

To address these limitations, the Wuppertal-Budapest lattice QCD collaboration developed a resummation scheme, henceforth referred to as $T'$-expansion scheme, that can reach high density with fewer terms in the expansion, exhibits smoother behavior at finite density, and effectively handles the expansion dependence on the QCD transition temperature~\cite{Borsanyi:2021sxv,Borsanyi:2022qlh}. This scheme has been used to develop an equation of state with 3D Ising critical point in Ref.~\cite{Kahangirwe:2024cny}. 
It is, however, crucial to thoroughly assess and explore the performance and limitations of this scheme. This involves comparing different truncation orders to the full result, to check how many terms are needed in each scheme to provide a good description of the full curve, discussing the schemes' radius of convergence, and comparing the behavior of the (truncated) observables at the same temperatures and chemical potentials, to see e.g. whether the unphysical wiggles that appear in some quantities as a truncation of the Taylor series are also present in the new expansion scheme.
To achieve this, we analyze the behavior of the  $T'$-expansion scheme when applied to effective QCD models designed for two different temperature regimes, namely the Hadron Resonance Gas~(HRG) model at low temperatures, and the Cluster Expansion Model~(CEM) at high temperatures.

The motivation for choosing these models is the following. The Hadron Resonance Gas is a very common model used to study, among other things, the thermodynamics of strongly interacting matter below the phase transition. Given its simple structure, and given the fact that it is often used to extend lattice QCD thermodynamics to low temperatures, it is important to see how the different expansion schemes work for it. 
As for the CEM, we chose it because it describes the thermodynamics around the phase transition. 
There, the Taylor coefficients show a characteristic wiggly structure that is reflected in the Taylor-truncated thermodynamics. It is therefore a good candidate to demonstrate the superiority of the $ T^\prime-$ expansion scheme. We also tested the expansion scheme on the Holographic model \cite{Hippert:2023bel} whose higher-order susceptibilities agree with lattice at zero chemical potential but did not include the results since the result is essentially the same as in CEM.
A similar analysis of the $T'$-expansion scheme was performed recently in Ref. \cite{Wen:2024hbz} in the context of the functional renormalization group approach.

This paper is organized as follows: in Section \ref{expansion}  we discuss the Taylor expansion and the two versions of the $T'$-expansion scheme, wherein we compute higher-order expansion coefficients. In Section \ref{EffectiveQCD} we briefly review the HRG and CEM effective frameworks.
We present the analysis of Taylor vs. $T'$-expansion schemes within these models in Section \ref{results} and conclude with the Summary in Section \ref{Summary}.

\section{\label{expansion}Expansion schemes}
\subsection{\label{expansionT}Taylor Expansion}

In the Taylor expansion, the pressure is expressed in Eq: \eqref{PressureTaylor} as a sum of all pressure derivatives, calculated on the lattice at $\mu_B=0$, multiplied by increasing powers of $\left(\frac{\mu_B}{T}\right)$, which serves as the expansion variable. 
Due to the charge conjugation symmetry, only even powers of $\mu_B/T$ contribute to the pressure
\begin{equation}
    \frac{P(T,\mu_B)}{T^4} = \sum_0^\infty
    \frac{\chi_{2n}^B(T,\mu_B = 0)}{2n!}
    \left(\frac{\mu_B}{T}\right)^{2n}, 
    \label{PressureTaylor}
\end{equation}
where the coefficients are baryon number susceptibilities, generically defined as
\begin{equation*}
    \chi_n^B(T,\mu_B) = 
    \left( \frac{\partial^n (P/T^4)}{\partial (\mu_B/T)^n} \right)_T.
\end{equation*} 

The baryon density is the first derivative of the pressure with respect to the chemical potential $\mu_B$. Its Taylor expansion reads
\begin{eqnarray}
    \frac{n_B(T,\mu_B)}{T^3} 
    =\sum_{n=1}^\infty 
    \frac{\chi_{2n}(T,\mu_B=0)}{(2n-1)!}
    \left(\frac{\mu_B}{T}\right)^{2n-1}.
    \label{BaryonTaylor}
\end{eqnarray}

All other thermodynamic observables like energy density, entropy density etc. can be computed from the pressure, through thermodynamic relations.

\subsection{\label{sec:citeref}$T'$ expansion scheme}

Another extrapolation scheme is the $T'-$expansion scheme introduced by the Wuppertal-Budapest collaboration in Ref. \cite{Borsanyi:2021sxv}.
The scheme was developed to address the shortcomings of a Taylor expansion truncated at low order, due to the lack of lattice QCD results on high-order susceptibilities.
This expansion has originally been based on the observation from lattice simulations at imaginary baryochemical potential that the normalized baryon density $T \chi_1^B/\mu_B$ at non-zero imaginary $\mu_B$ approximately coincides in the transition region with the second order susceptibility $\chi_2^B$ at $\mu_B = 0$ evaluated at a shifted temperature, $T + \kappa_2 \mu_B^2 / T$.
This shifting behavior is predominant in the vicinity of the transition line, where the slope of the scaled baryon density exhibits larger variations compared to the data at low and high temperatures. 

One can define the following systematic scheme for mapping $T \chi_1^B/\mu_B$ at finite $\mu_B$ into $\chi_2^B$ at $\mu_B = 0$:
\begin{equation}
\frac{\chi_1^B(T,\hat{\mu}_B)}{\hat{\mu}_B} = \chi_2^B(T',0) \label{altexsID}
\end{equation}
where the shifted temperature $T'$ is expressed as a series in $\hat{\mu}_B = \mu_B/T$:
\begin{equation}
    T'(T,\hat{\mu}_B) = T\left[1+ \kappa_2^B(T)\hat{\mu}_B^2 + \kappa_4^B(T)\hat{\mu}_B^4 + \kappa_6^B(T)\hat{\mu}_B^6 + ...\right]. 
    \label{Eq:Tprime}
\end{equation}
Equations~\eqref{altexsID} and \eqref{Eq:Tprime} define the $T'-$expansion scheme, which can be viewed as a particular resummation of the Taylor expansion.
By matching the baryon density in Taylor and $T'-$expansions order-by-order in $\hat{\mu}_B$, one can define the coefficients $\kappa_n^B$ in terms of the susceptibilities $\chi_n^B$~\cite{Borsanyi:2021sxv}:
\begin{align} \nonumber
    \kappa_2^{B} &= \frac{1}{T\chi_2'^B} \left(\frac{\chi_4^B}{3!}\right) \\ \nonumber
    \kappa_4^{B} &= \frac{1}{T\chi_2'^{B}}\left(\frac{\chi_6^B}{5!} - \frac{T^2}{2!}\chi_2''^B\kappa_2^{B^2}\right)  \\
    \kappa_6^{B} &= \frac{1}{T\chi_2'^{B}}\left(\frac{\chi_8^B}{7!} - \frac{T^2}{2!}\chi_2^{B''}(2\kappa_2^{B}\kappa_4^{B}) - \frac{T^3}{3!}\chi_2^{B'''}\kappa_2^{B^3}\right) ~.
    \label{Eqkappas}
\end{align}

\subsection{$T'$ expansion scheme with the Stefan-Boltzmann limit constraint}

It has been observed that, in the strangeness neutral case, the previously developed $T'$-expansion scheme 
does not perform well at high temperatures~\cite{Borsanyi:2022qlh}. 
This scheme, with different lines collapsing at $\mu_B=0$ when shifted by a constant $\kappa_2$, was expected to falter because the approximate scaling variable falls outside the crossover range. 
This limits its applicability in the high-temperature regime, due to the restriction imposed by the Stefan-Boltzmann limit of free quarks. 
To ensure that the main identity holds even as $T \rightarrow\infty$, in \cite{Borsanyi:2022qlh} a generalization was suggested by normalizing the thermodynamic quantities by their the Stefan-Boltzmann limits.
This adjustment enhances convergence at high temperatures.
The Stefan-Boltzmann limit compatible $T'-$expansion scheme reads 
\begin{align}
n_B(T,\mu_B) = \frac{\bar{\chi}_1^B(\hat{\mu}_B)}{\bar{\chi}_2^B(0)}\chi_2^B(T'(T,\mu_B),0)
\label{Eq:TprimenB_SB}
\end{align}
where  $\bar{\chi}_n^B(\hat{\mu}_B)$ are baryon susceptibilities in the Stefan-Boltzmann limit and
\begin{align}
\label{eq:TprimeSB}
T'(T,\mu_B) = T\left(1+\lambda_2^B(T)\hat{\mu}_B^2 + \lambda_4^B(T)\hat{\mu}_B^4 + \lambda_6^B(T)\hat{\mu}_B^6 + ...\right).
\end{align}
The new expansion coefficients are denoted by $\lambda_n^B$. 
They are expressed in terms of the susceptibilities in the same way as before, by matching to the Taylor expansion order-by-order in $\mu_B/T$:
\begin{align}
\nonumber
    \lambda_2^{B} &= \frac{1}{T\chi_2'^B} \left(\frac{\chi_4^B}{3!}-\frac{c_4}{c_2}\chi_2^B\right) \\ 
    \nonumber
    \lambda_4^{B} &= \frac{1}{T\chi_2'^{B}}\left(\frac{\chi_6^B}{5!} - \frac{c_4}{c_2}\lambda_2 T \chi_2^{B '}  -\frac{T^2}{2!} \lambda_2^2  \chi_2^{B ''}\right) \\ 
    \lambda_6^{B} &= \frac{1}{T\chi_2'^{B}}\left(\frac{\chi_8^B}{7!} - \frac{c_4}{c_2}\lambda_4 T \chi_2^{B '}  -\frac{T^2}{2!}\frac{c_4}{c_2} \lambda_2 \lambda_4 \chi_2^{B ''} - \frac{T^3}{3!} \lambda_2^3  \chi_2^{B'''}\right).
    \label{Eq:lambda}
\end{align}
Here 
$c_2 = \frac{1}{3},$ $c_4 = \frac{1}{27\pi^2}$, and $\bar{\chi}_1^B(\hat{\mu}_B) = c_2 \hat{\mu}_B + c_4 \hat{\mu}_B^3$.

\section{\label{EffectiveQCD}{Effective QCD models}}

\subsection{\label{HRG}Hadron Resonance Gas Model}

The Hadron Resonance Gas~(HRG) Model describes the hadronic phase of QCD as a multi-component system of hadrons and resonances, reflecting the concept of resonance dominance of interactions following the early ideas of Rolf Hagedorn~\cite{Hagedorn:1968jf}.
The list of hadrons and resonances is usually taken from Particle Data Group (PDG) listings~\cite{10.1093/ptep/ptac097}. 
The HRG model
has shown remarkable agreement with the state-of-the-art lattice QCD results for QCD thermodynamics at temperatures below $T_{\rm pc} \simeq 155$ MeV.
In this section with briefly discuss properties of three variants of the HRG model: (i) the Ideal HRG model that corresponds to a non-interacting gas of hadrons and resonances, (ii) the Excluded-Volume HRG~(EV-HRG) model that includes repulsive  core~\cite{Vovchenko:2017xad}, and (iii) the van der Waals HRG~(vdW-HRG) model which includes both attractive and repulsive interactions and the nuclear liquid-gas transition~\cite{Vovchenko:2016rkn}.   

We neglect quantum statistics and consider the $\mu_S=0,~\mu_Q=0$ case. 
We incorporate EV/vdW interactions for pairs of baryons and pairs of antibaryons only.
We use the particle list PDG2021+~\cite{SanMartin:2023zhv} which does not include light nuclei. 

\subsubsection{Ideal HRG}

In the Ideal HRG model, the baryon number density is given by
\begin{align}
n_B(T,\mu_B) = 2 \phi_B(T) \sinh(\mu_B/T),
\label{eq:nBIG}
\end{align}
where
\begin{align}
\phi_B(T) & = \sum_{i\in B}\frac{d_im_i^2T}{2\pi^2}  K_2(m_i/T),
\label{phi}
\end{align}
where $d_i$ and $m_i$ represent the degeneracy and mass of particle $i$, respectively, and  $K_2$ is the modified Bessel function of the second kind. 
All even order baryon number susceptibilities are equal in the ideal HRG model, $\chi_{2n}^B (T) = 2 \phi_B(T)$.
 We use \texttt{Thermal-FIST}~\cite{Vovchenko:2019pjl} to calculate $\phi_B(T)$.

\subsubsection{EV-HRG}

In the EV-HRG with repulsive baryon core, the expression for baryon number density is given in terms of the Lambert $W$ function~\cite{Noronha-Hostler:2012ycm,Taradiy:2019taz}
\begin{align}
n_B(T,\mu_B) = \frac{1}{b} \left\{ \frac{W(x_+)}{1 + W(x_+)} - \frac{W(x_-)}{1 + W(x_-)}  \right\},
\label{eq:nBEV}
\end{align}
where $x_\pm = b \phi_B(T) e^{\pm \mu_B/T}$.

The EV-HRG model reduces to the Ideal case in the limit $b \to 0$.
In our EV-HRG model calculations, we take $b=1~\text{fm}^3$, as motivated by fits to lattice QCD data on baryon number susceptibilities~\cite{Karthein:2021cmb}.

\subsubsection{vdW-HRG}

In the case of the vdW-HRG model, the densities of baryons ($n_{B^+}$) and antibaryons ($n_{B^-}$) are determined by the following pair of independent transcendental equations:
\begin{align}
      b \phi_B(T) e^{\pm\mu_B/T} & = \frac{b n_{B^+(B^-)}}{1-bn_{B^+(B^-)}} \nonumber \\
      & \quad \times \exp\left[\frac{bn_{B^+(B^-)}}{1-bn_{B^+(B^-)}} - \frac{2an_{B^+(B^-)}}{T}\right].
    \label{tran1}
\end{align}
Parameters $a$ and $b$ correspond to attractive and repulsive baryon interactions, 
respectively. 
Once the numerical solution for the densities $n_{B^+}$ and $n_{B^-}$ is obtained, baryon number susceptibilities can be computed analytically to the desired order via an iterative procedure based on the Fa\`a di Bruno formula~\cite{Savchuk:2019yxl}.
We use this to compute susceptibilities up to $\chi_8^B$, which are necessary to calculate $\kappa_6^B$ and $\lambda_6^B$.

For $a = 0$, the model reduces to the EV-HRG, where an explicit solution of Eq.~\eqref{tran1} is given in terms of the Lambert $W$ function~\eqref{eq:nBEV}.
For $a = 0$ and $b = 0$, one obtains the ideal HRG model.
In the case of vdW-HRG model, these parameters are assigned values of $a=329  ~\text{MeV} ~ \text{fm}^3$ and $b=3.42~\text{fm}^3$ to reproduce the properties of the nuclear liquid-gas transition, as derived in \cite{Vovchenko:2015vxa,Sanzhur:2022jxx}. 

\subsection{\label{CE}{Cluster Expansion Model}}

In addition to the HRG model at low temperatures, we utilize the Cluster Expansion Model~(CEM)~\cite{Vovchenko:2017gkg,Vovchenko:2018zgt} to explore the high-temperature regime, particularly near and above the QCD phase transition. 
The CEM approach makes use of the fugacity expansion and explicitly preserves the Roberge-Weiss periodicity of the QCD partition function $Z$, namely $Z(\mu_B) = Z(\mu_B + i2\pi T)$~\cite{Roberge:1986mm}.
 
The relativistic fugacity expansion for the baryon density reads
\begin{align}
\label{eq:nBFug}
    n_B(T,\mu_B)=\sum_{k=1}^\infty b_k(T)\sinh\left(\frac{k\mu_B}{T}\right).
\end{align} 
Here, $b_k(T)$ are the Fourier coefficients of the baryon number density, which can be computed through lattice QCD simulations at imaginary chemical potential~\cite{Vovchenko:2017xad,Bellwied:2021nrt}.
They are related to baryon number susceptibilities through
\begin{align}
\chi_{2n}^B(T) = \sum_{k=1}^\infty k^{2n-1}b_k(T).
\label{Eq:chi_CEM}
\end{align} 

The leading four Fourier coefficients have been computed on the lattice in Ref.~\cite{Vovchenko:2017xad}.
For $T > 160$~MeV, the lattice data for $b_3(T)$ and $b_4(T)$ are consistent with the Stefan-Boltzmann-limit-based scaling
\begin{align}
\label{eq:CEMscale}
b_k(T) = \frac{[\hat{b}_2(T)]^{k-1}}{[\hat{b}_1(T)]^{k-2}} b_k^{\rm SB}, \quad k = 3,4, \ldots
\end{align} 
where $\hat{b}_{k} = \frac{b_{k}(T)}{b_{k}^{SB}}$ and $b_k^{SB} = \frac{(-1)^{k-1}}{k}\frac{4[3+4(\pi k)^2]}{27(\pi k)^2}$ are the Fourier coefficients in the Stefan-Boltzmann limit.
The CEM assumes that the scaling~\eqref{eq:CEMscale} holds for all coefficients $k \geq 3$.
In this case, the fugacity expansion~\eqref{eq:nBFug} for baryon number density can be summed analytically~\cite{Vovchenko:2018zgt}, giving
\begin{align}\nonumber
    \frac{n_B(T,\mu_B)}{T^3} &= -\frac{2}{27\pi^2}\frac{\hat{b}_1^2}{\hat{b}_2}\{4\pi^2[\operatorname{Li}_1(x_+) - \operatorname{Li}_1(x_{-}  )] \\ 
    & \quad +  3[\operatorname{Li}_3(x_+) -\operatorname{Li}_3(x_-)]\},  \label{Eq:nBCEM}
\end{align}
where 
$x_\pm = -\frac{\hat{b}_2}{\hat{b}_1}e^{\pm \mu_B/T}$ and $\operatorname{Li}_s(z) = \sum_{k=1}^\infty \frac{z^k}{k^s}$ is the polylogarithm.
Baryon number susceptibilities in the CEM read
\begin{eqnarray}
    \chi_k^B(T,\mu_B) &=& -\frac{2}{27 \pi^2} \frac{\hat{b}_1^2}{\hat{b}_2} \left\{ 4\pi^2 \left[ \text{Li}_{2-k}(x_{+}) + (-1)^k \text{Li}_{2-k}(x_{-}) \right]  \right.\nonumber \\
    && \left. + 3 \left[ \text{Li}_{4-k}(x_{+}) + (-1)^k \text{Li}_{4-k}(x_{-}) \right] \right\}.
\end{eqnarray}
Besides the Taylor expansion coefficients, we use the above expression to compute the coefficients  $\kappa_n^B$  in Eq. \eqref{Eqkappas} and $\lambda_n^B$  in Eq. \eqref{Eq:lambda} for the two $T'$-expansion schemes under consideration.

The CEM uses lattice QCD on leading Fourier coefficients $b_1(T)$ and $b_2(T)$ of baryon density as input and describes well the resulting high-order baryon number susceptibilities~\cite{Vovchenko:2017gkg}.
The CEM provides a lattice-based equation of state model without a critical point. The CEM can be improved by incorporating errors in the lattice inputs.
Here it is not essential for the purposes of our studies of convergence properties of the CEM, therefore we do not incorporate lattice errors.

\subsection{Taylor expansion convergence radius}
The radius of convergence for a Taylor expansion in $\mu_B/T$
corresponds to the closest singularity in the partition function in the complex $\mu_B/T$ plane.
Physically, these singularities may correspond to the QCD critical point~\cite{Stephanov:2006dn}, Roberge-Weiss transition~\cite{Oshima:2023bip}, nuclear liquid-gas transition~\cite{Savchuk:2019yxl}, or thermal singularities in Fermi-Dirac/Bose-Einstein distribution functions~\cite{Skokov:2010uc}.
At finite volume, these correspond to Lee-Yang zeros that are used in lattice-based studies~\cite{Connelly:2020pno,Basar:2021gyi, Clarke:2024ugt}. 
For the models under consideration, the radius of convergence can be determined explicitly.

\subsubsection{HRG models}

In the Ideal HRG model, the baryon density reads $n_B = 2 \phi_B(T) \sinh(\mu_B/T)$ and the radius of convergence is infinite
\footnote{If quantum statistics were included, the radius of convergence would correspond to the zero of the inverse Fermi-Dirac function for nucleons, giving $r_{\mu_B/T} = m_N/T$~\cite{Taradiy:2019taz}.}.
 
The presence of repulsive interactions in the EV-HRG model leads to a finite radius of convergence, determined by the branch cut singularity of the Lambert $W$ function in Eq.~\eqref{eq:nBEV} at $W[b \phi_B(T) e^{\pm \mu_B/T} = -e^{-1}]$.
The radius of convergence reads~\cite{Taradiy:2019taz}
\begin{align}
    r_{\mu_B/T} = \sqrt{\{1+ \ln[b\phi_B(T)]\}^2+\pi^2}.
    \label{Eq:radiusEV}
\end{align}

In the vdW-HRG model, the singularity which limits the convergence radius of the Taylor expansion is connected to the spinodal points of the nuclear liquid-gas transition, which are located in the complex $\mu_B/T$ plane for temperatures $T > T_c = \frac{8a}{27b} \simeq 28.5$~MeV\footnote{With quantum statistics included, the critical temperature is lower, $T_c \simeq 19.7$~MeV~\cite{Vovchenko:2015vxa}.}.
They are determined as solution to the following equation
\begin{align}
\frac{d\mu_B}{d n_{B^+}}\bigg|_{\mu_B = \mu_B^{\rm br}} = 0,
\label{Eq:dmuBdnB}
\end{align}
with $\mu_B^{br}$ being the location of the limiting singularity. 
The equation for branch points can be obtained explicitly by differentiating Eq.~\eqref{tran1}, giving~\cite{Savchuk:2019yxl}
\begin{align}
\frac{2a n_{\rm br}}{T} (1 - b n_{\rm br})^2 = 1.
\end{align}
This is a cubic equation for $n_{\rm br}$, which can be solved explicitly.
Plugging the resulting value of $n_{\rm br}$ into Eq.~\eqref{tran1} allows one to obtain $\mu_B^{\rm br}$.
The radius of convergence then reads
\begin{align}
   r_{\mu_B/T} = |\mu_B^{\rm br}/T| = \frac{\sqrt{[{\rm Re}(\mu_B^{\rm br})]^2 + [{\rm Im}(\mu_B^{\rm br})]^2}}{T}.
   \label{Eq:rmuBT}
\end{align}

\subsubsection{Cluster Expansion Model}
The radius of convergence of the Taylor expansion for the CEM is determined by the branch cut singularity of the polylogarithm, $\operatorname{Li_k}(x_\pm = 1)$, giving
\begin{align}
\label{eq:rmuTCEM}
    r_{\mu_B/T} = \sqrt{[\ln(-\hat{b}_1/\hat{b}_2)]^2 + \pi^2}.
\end{align}

We then use susceptibilities from Eq. \eqref{Eq:chi_CEM} to compute the $\kappa_n^B$'s in Eq. \eqref{Eqkappas} for the $T'-$expansion, and the $\lambda$s in Eq. \eqref{Eq:lambda} for the $T'-$expansion with the Stefan-Boltzmann limit.

\section{\label{results}{Results}}

We test all three expansion schemes, Taylor, $T'$-expansion, and $T'$-expansion corrected for Stefan-Boltzmann limit, on the Ideal-, EV-, and vdW-HRG models, as well as on the Cluster Expansion Model.

\subsection{Hadron Resonance Gas model} 

\subsubsection{Taylor Expansion}
\begin{figure}[h!]
    \includegraphics[scale=0.275]{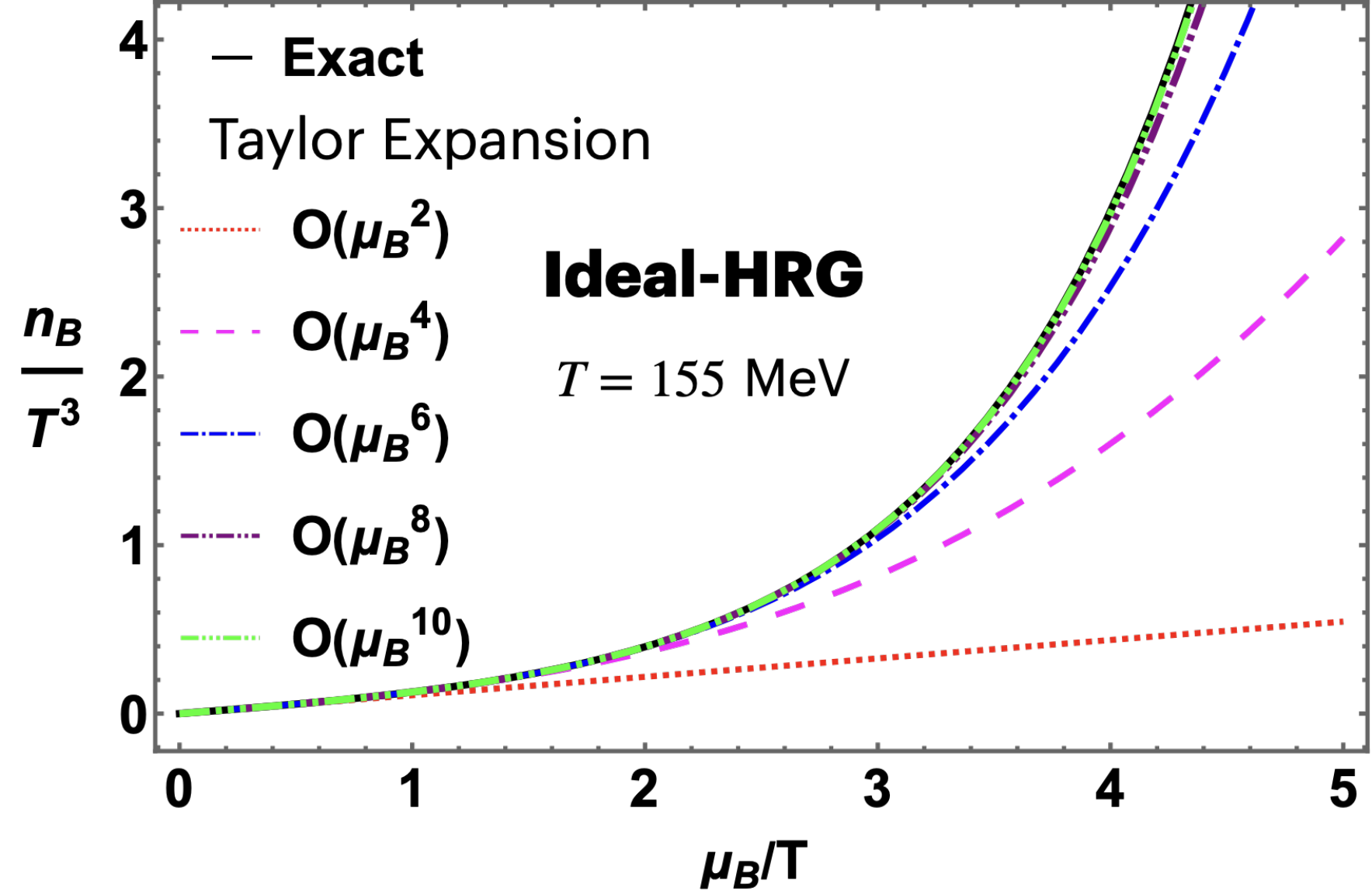}\\
\includegraphics[scale=0.275]{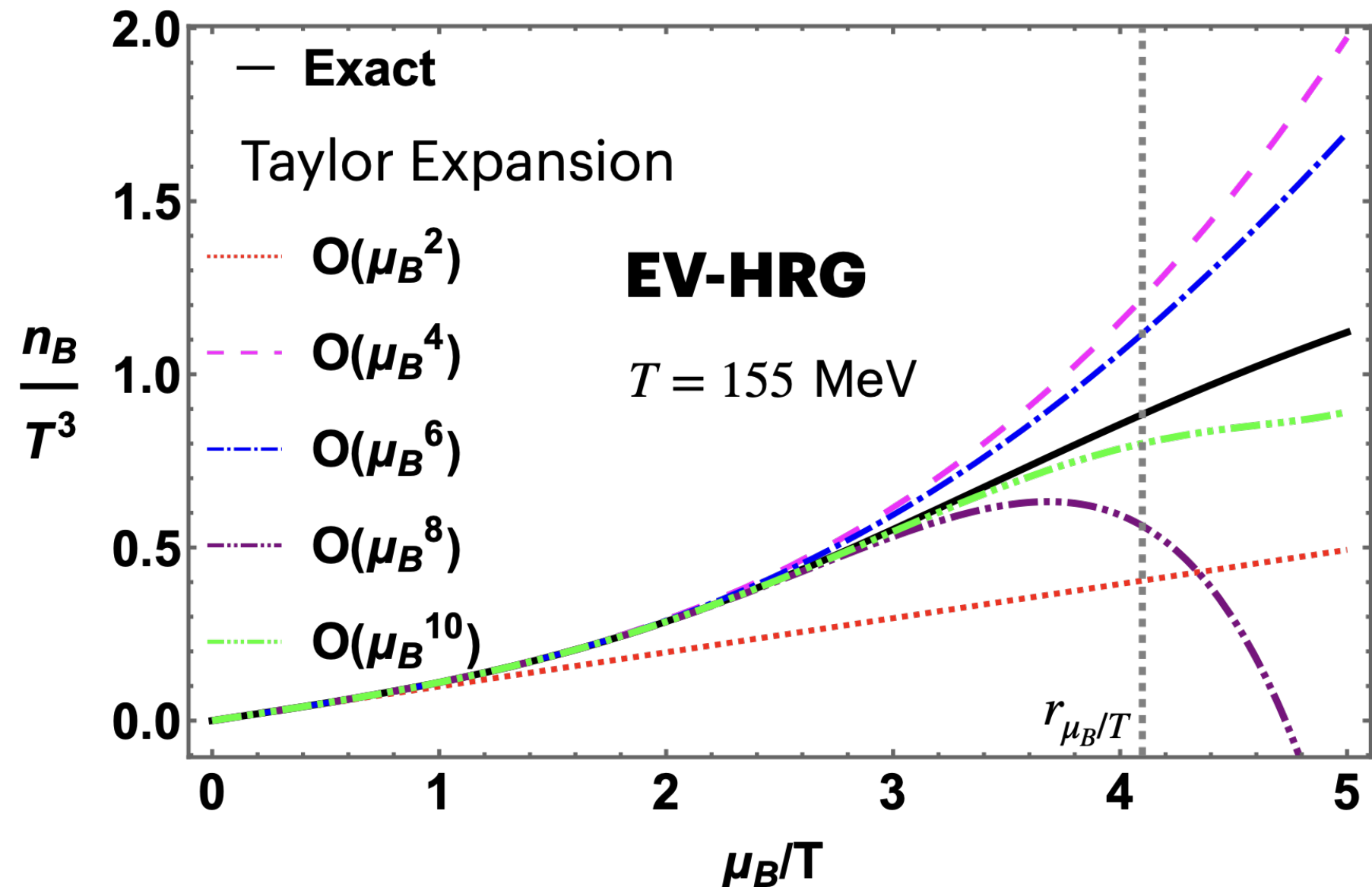}
    \centering
\includegraphics[scale=0.245]{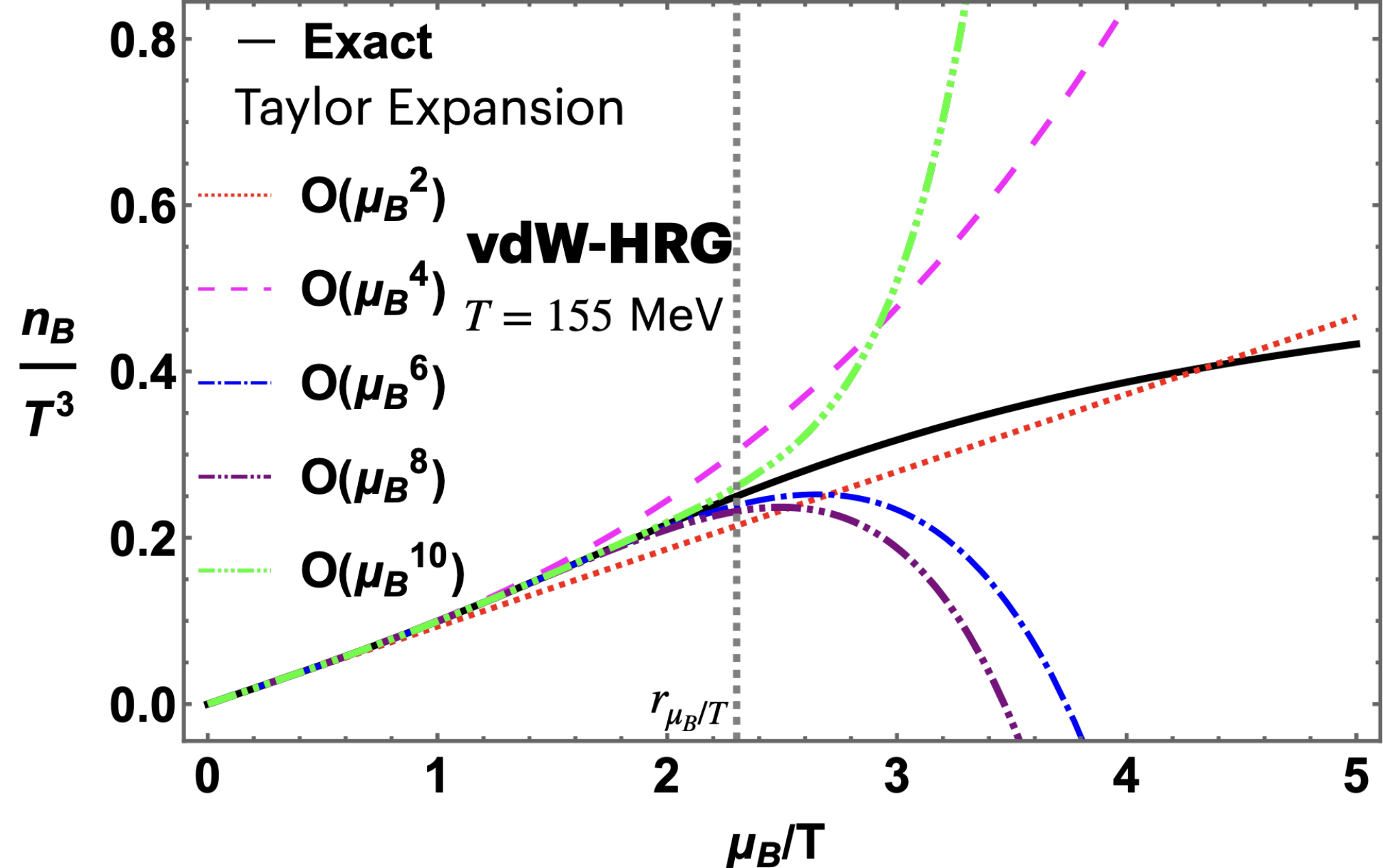}
   \caption{Scaled baryon density $n_B/T^3$ as a function of \(\mu_B/T\) from different HRG models at \(T = 155~\, \text{MeV}\): Ideal (top panel), EV-HRG (center panel), vdW-HRG (lower panel). In all panels, the solid black line represents the exact result, while the colored dashed lines indicate different orders of the Taylor expansion truncation. The vertical dotted black lines in the center and bottom panels denote the radius of convergence for the Taylor expansion, corresponding to EV-HRG and vdW-HRG models, obtained using Eq. \eqref{Eq:radiusEV}
 and Eq. \eqref{Eq:rmuBT}, respectively.}
    \label{ordersTaylor}
\end{figure}
Figure \ref{ordersTaylor} shows the Taylor expansion tested on the three versions of the HRG model: Ideal (top panel), Excluded Volume (center panel), van der Waals (bottom panel). In all cases, $n_B/T^3$ is plotted as a function of $\mu_B/T$, at a temperature $T=155$ MeV. In all panels, the exact result is shown as a solid black curve, while the other curves correspond to different truncation orders of the Taylor expansion: $\mathcal{O}(\mu_B^2)$ (red, dotted), $\mathcal{O}(\mu_B^4)$ (magenta, dashed), $\mathcal{O}(\mu_B^6)$ (blue, dot-dashed), $\mathcal{O}(\mu_B^8)$ (purple, double dot-dashed), $\mathcal{O}(\mu_B^{10})$ (green, triple dot-dashed). 

For the Taylor expansion case, as expected for the Ideal HRG, the radius of convergence is infinite, as illustrated in the top plot in Fig. \ref{ordersTaylor}. In contrast, both the EV-HRG and vdW-HRG exhibit a finite radius of convergence, indicated as a vertical dotted line in the center and bottom panels. In these models, the different orders of expansion converge within the radius of convergence and deviate from the exact solution beyond this value of $\mu_B/T$. In all cases, higher order truncations show a better agreement with the full result within the radius of convergence, and represent a good approximation of the full result up to larger values of $\mu_B/T$.

\subsubsection{$T'$-expansion scheme}
\begin{figure}[t!]
    \centering
   \includegraphics[scale=0.27]{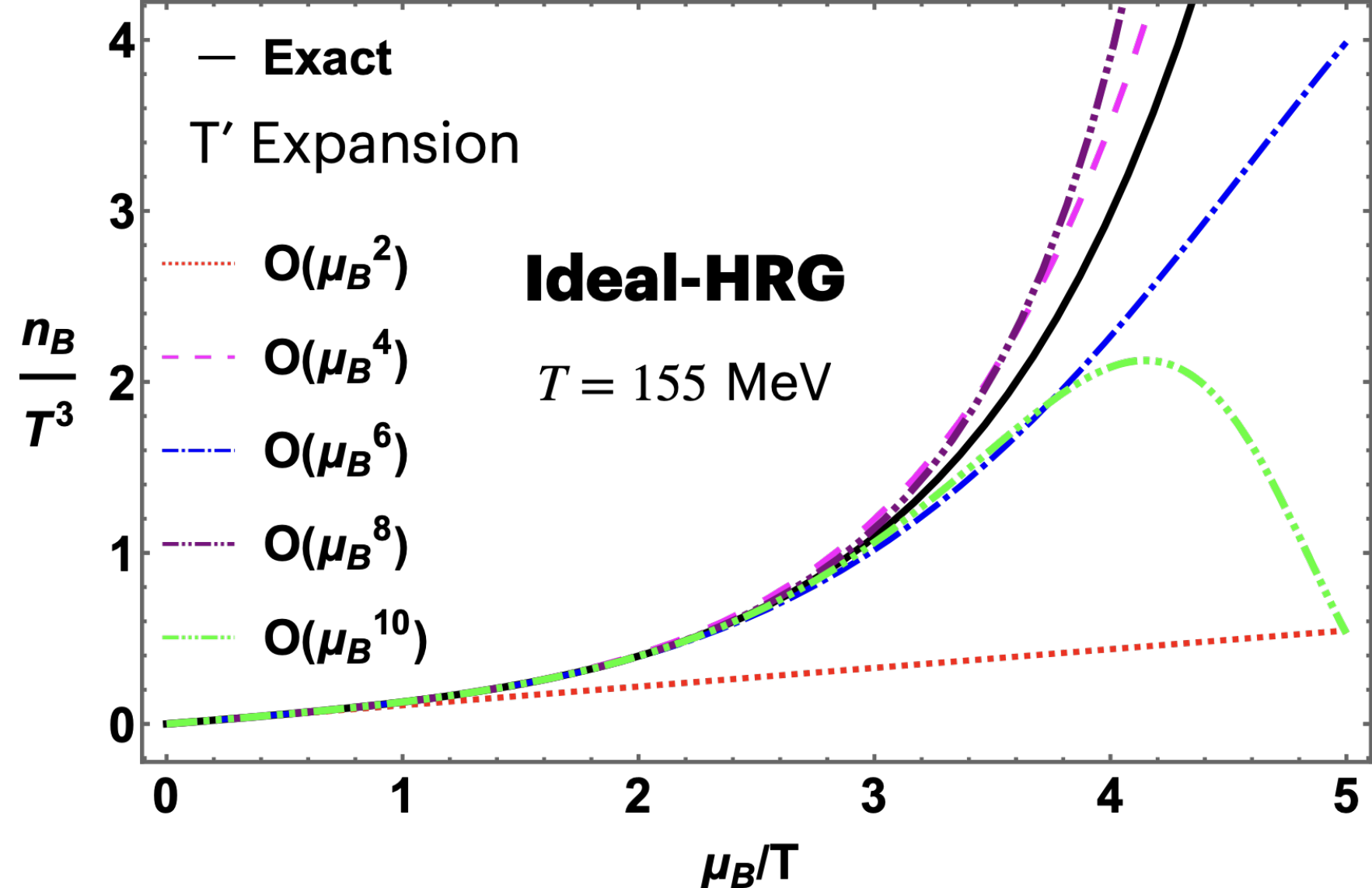}\\
    \centering
\includegraphics[scale=0.26]{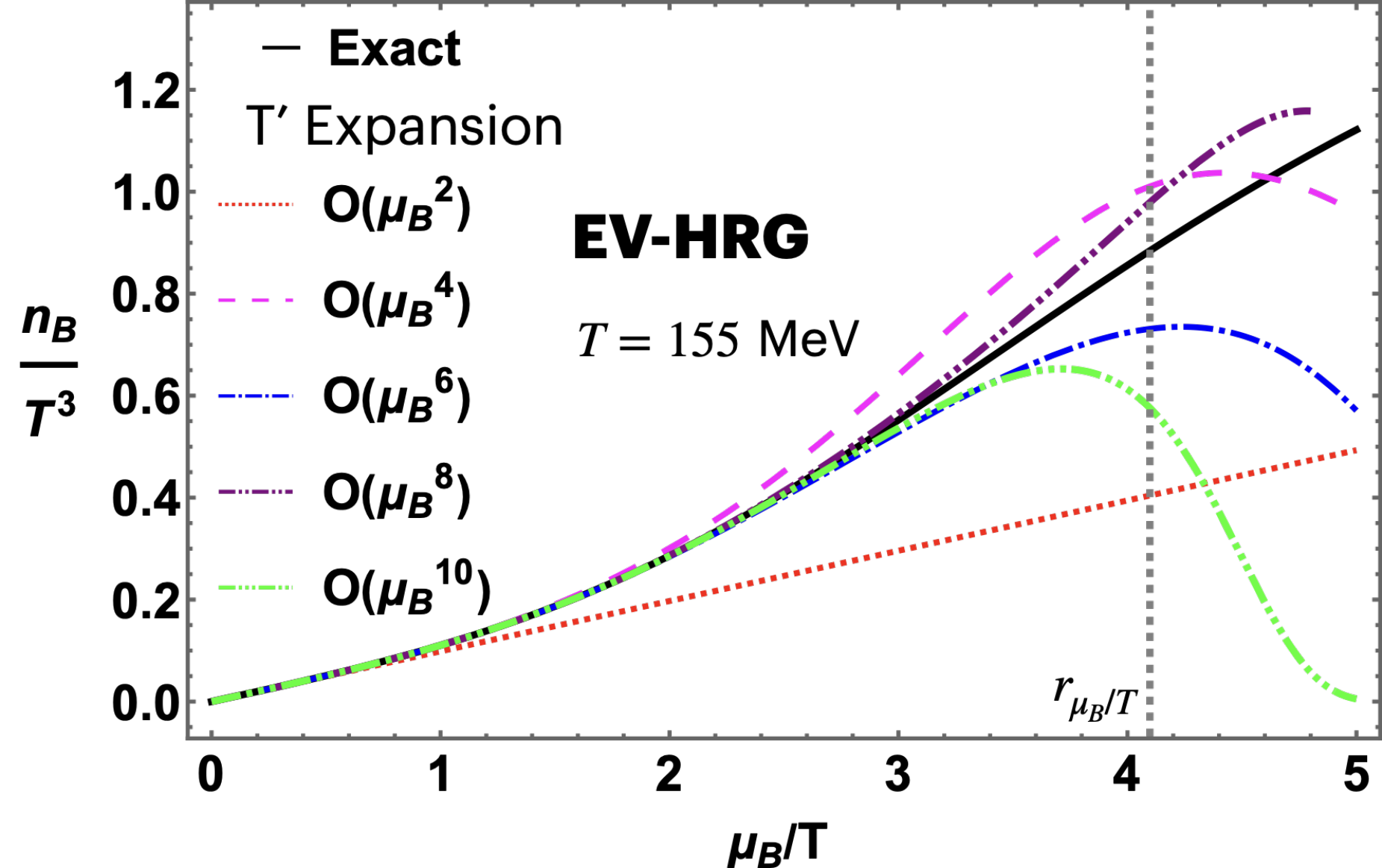}\\
    \centering
    \includegraphics[scale=0.28]{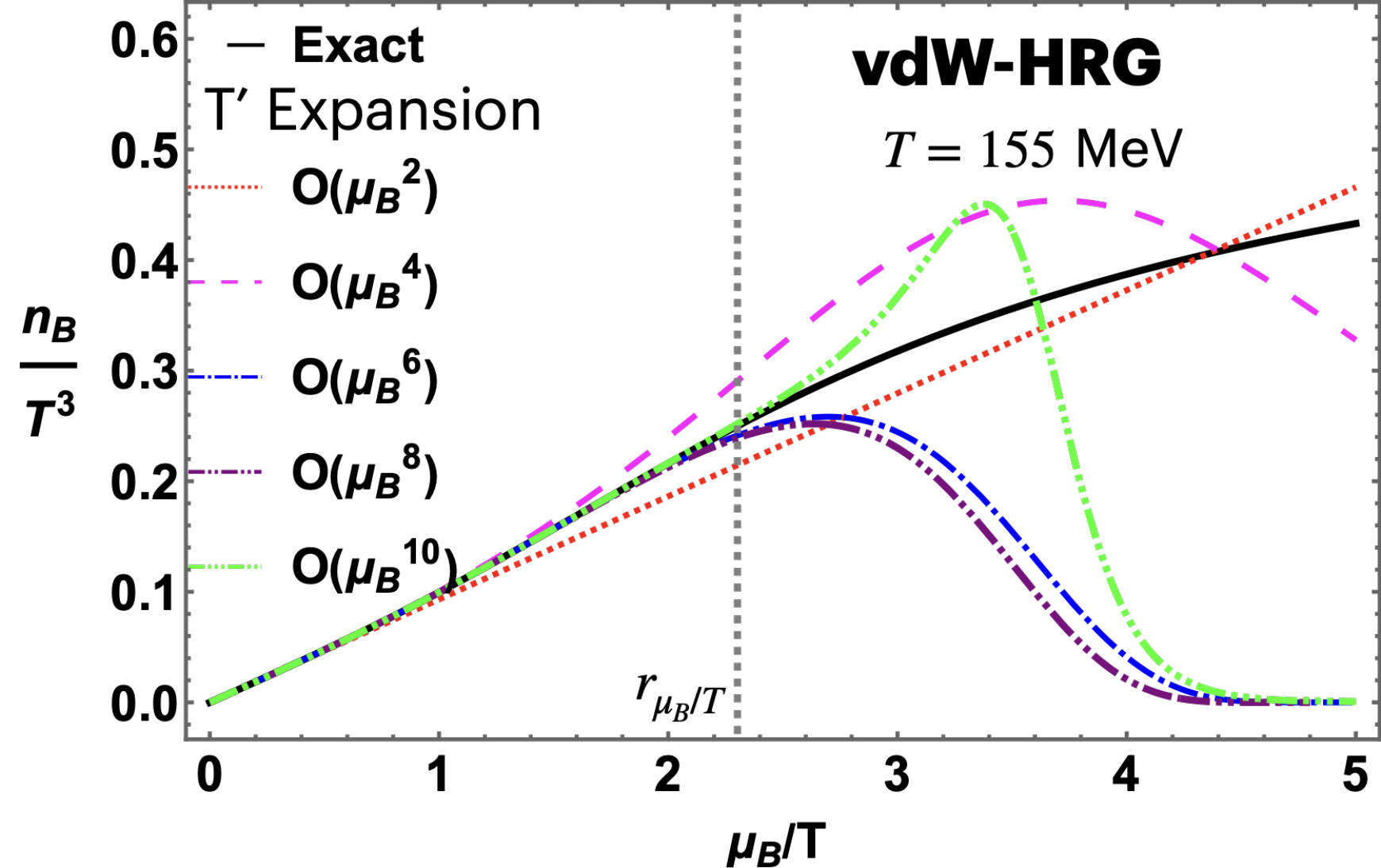} 
    \caption{The three panels show the scaled baryon density $n_B/T^3 $ as a function of $\mu_B/T$ from different HRG models at $T=155 ~\text{MeV}$: Ideal (top panel), Excluded Volume (center panel), van der Waals (bottom panel). In all panels, the solid black line represents the exact result, while colored lines indicate different orders of $T'$-expansion scheme truncation. Vertical dotted black lines in the center and bottom plots denote the radius of convergence for the Taylor expansion corresponding to EV-HRG and vdW-HRG models, obtained using Eq. \eqref{Eq:radiusEV} and Eq. \eqref{Eq:rmuBT}, respectively.}
    \label{TExS}
\end{figure}

Figure \ref{TExS} shows the $T'$-expansion scheme tested on the three versions of the HRG model: Ideal (top panel), Excluded Volume (center panel), van der Waals (bottom panel). In all cases, $n_B/T^3$ is plotted as a function of $\mu_B/T$, at a temperature $T=155$ MeV. In all panels, the exact result is shown as a solid black curve, while the other curves correspond to different truncation orders of the $T'$-expansion scheme: $\mathcal{O}(\mu_B^2)$ (red, dotted), $\mathcal{O}(\mu_B^4)$ (magenta, dashed), $\mathcal{O}(\mu_B^6)$ (blue, dot-dashed), $\mathcal{O}(\mu_B^8)$ (purple, double dot-dashed), $\mathcal{O}(\mu_B^{10})$ (green, triple dot-dashed). 

It appears that the $T'-$expansion scheme performs similarly to Taylor: higher order results show a better agreement with the full result, compared to the lower order ones. Besides, curves start deviating from each other and from the full result beyond the Taylor radius of convergence, which is still indicated as a dotted vertical line in the two lower panesl of Fig. \ref{TExS}. However, we notice that, if we truncate the $T'-$expansion to low order in $\mu_B$, we are able to achieve an agreement with the full result up to higher values of $\mu_B/T$, compared to the Taylor series truncated to the same order. This is evident e.g. when comparing the distance between the dashed magenta line and the full black line in Figs. \ref{ordersTaylor} and \ref{TExS}.

This observation hints that the $T'$-expansion scheme behaves as an asymptotic series divergent beyond a certain convergence radius, hence its worsening performance with the inclusion of further terms.
We observe that $\mathcal{O}(\mu_B^4)$ seems to provide the optimal performance possible for the HRG class of models.

\newpage
\subsubsection{$T'$-expansion scheme with SB limit}
\begin{figure}[h!]
    \centering
 \includegraphics[scale=0.28]{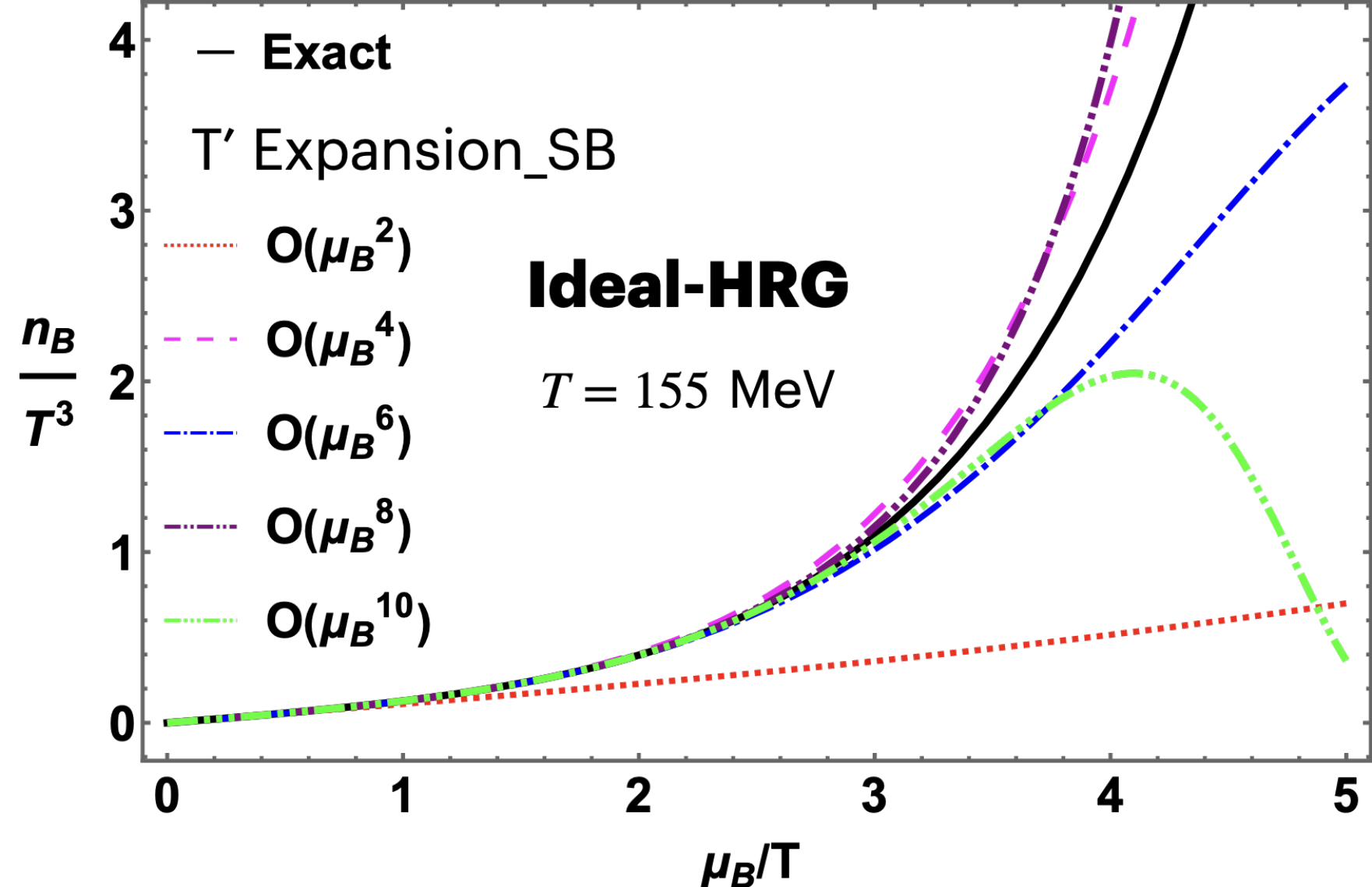}
\\
    \centering
    \includegraphics[scale=0.27]{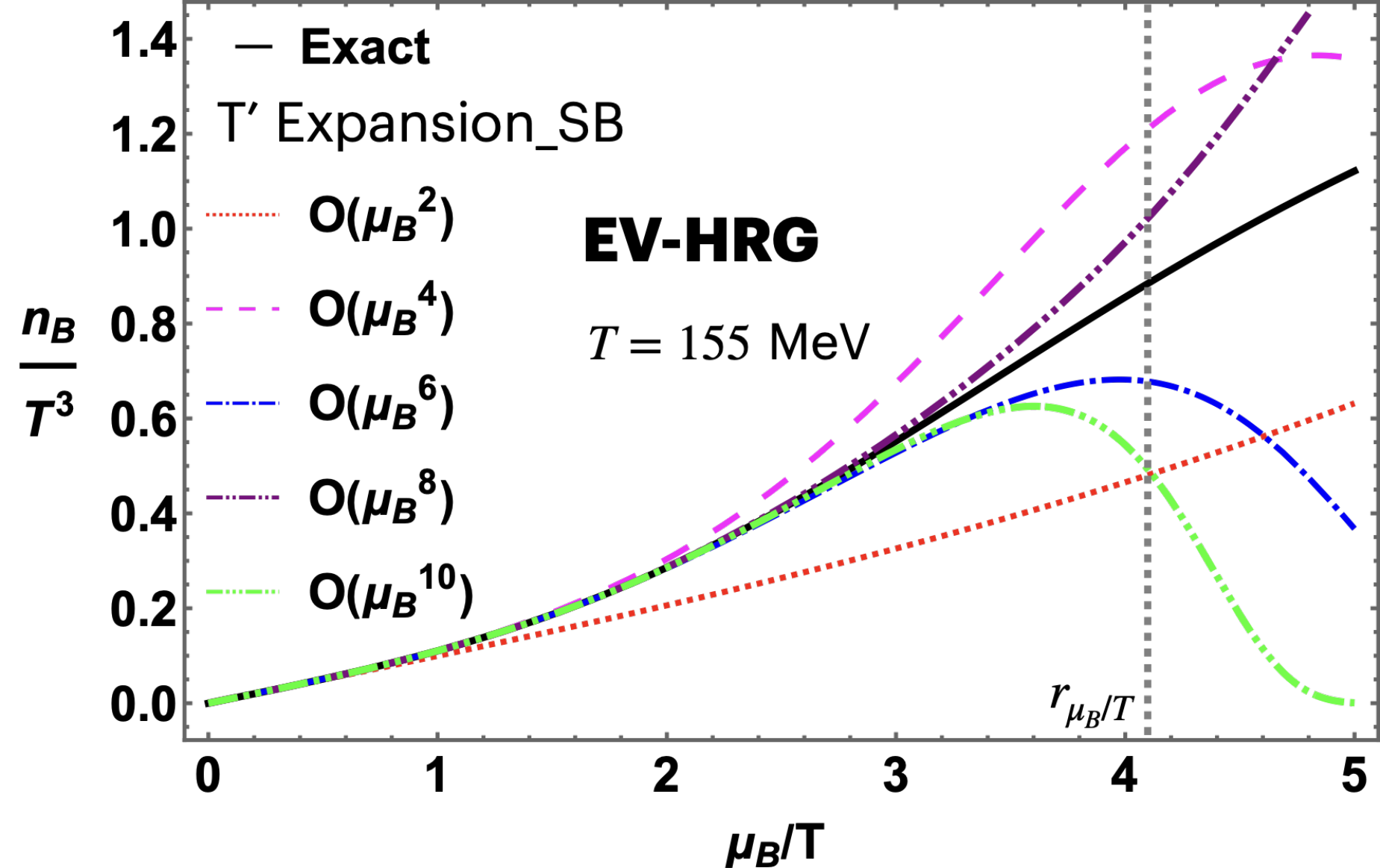}
\\
    \centering
    \includegraphics[scale=0.255]{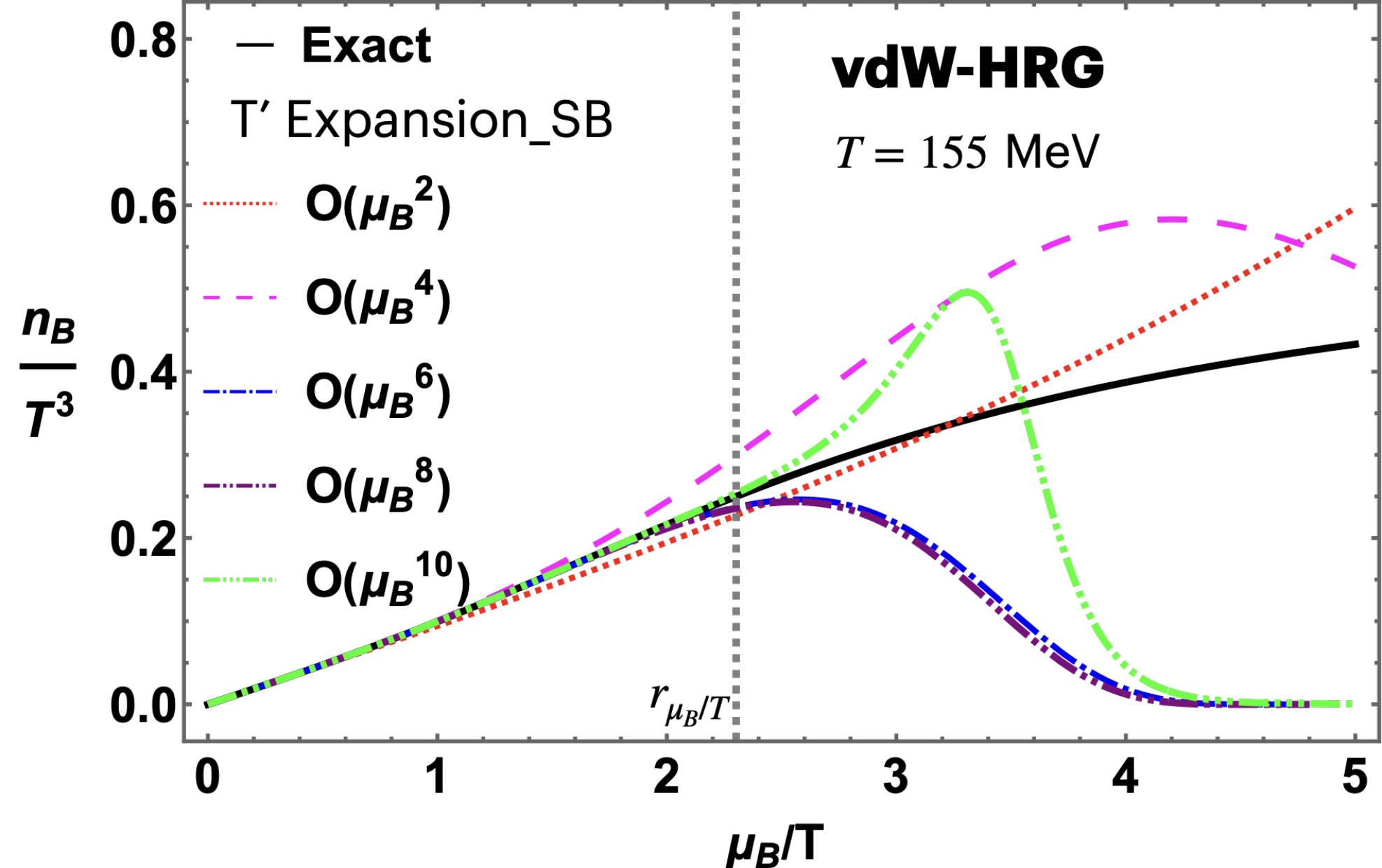} 
\caption{The three panels show the scaled baryon density $n_B/T^3 $ as a function of $\mu_B/T$ from different HRG models at $T=155~ \text{MeV}$: Ideal (top panel), Excluded Volume (center panel), van der Waals (bottom panel). In all panels, the solid black line represents the exact result, while colored lines indicate different orders of truncation of the $T'$-expansion scheme with SB limit. Vertical dotted black lines in the center and bottom plots denote the radius of convergence for the Taylor expansion corresponding to EV-HRG and vdW-HRG models, obtained using Eq. \eqref{Eq:radiusEV} and Eq. \eqref{Eq:rmuBT}, respectively. }
    \label{TExSSB}
\end{figure}

Figure \ref{TExSSB} shows the $T'$-expansion scheme, including the correct Stefan-Boltzmann limit, tested on the three versions of the HRG model: Ideal (top panel), Excluded Volume (center panel), van der Waals (bottom panel). In all cases, $n_B/T^3$ is plotted as a function of $\mu_B/T$, at a temperature $T=155$ MeV. In all panels, the exact result is shown as a solid black curve, while the other curves correspond to different truncation orders of the $T'$-expansion scheme with SB limit: $\mathcal{O}(\mu_B^2)$ (red, dotted), $\mathcal{O}(\mu_B^4)$ (magenta, dashed), $\mathcal{O}(\mu_B^6)$ (blue, dot-dashed), $\mathcal{O}(\mu_B^8)$ (purple, double dot-dashed), $\mathcal{O}(\mu_B^{10})$ (green, triple dot-dashed). 

From the figure, it is evident that the performance of the two versions of the $T'-$expansion scheme is very similar. 
Also in this case, the different orders get closer to the full result when including higher powers of $\mu_B/T$, with the lower-order truncations showing a slightly better agreement with the full result up to larger values of $\mu_B/T$, compared to the corresponding Taylor expansion.
Given that the HRG model itself does not reduce to the SB limit of free quarks at high temperatures, there is no a priori reason for the SB limit to improve the convergence. Thus, the absence of improvement of the $T'-$expansion with SB limit is unsurprising.

At this point, we point out that no real advantage of either version of the $T'-$expansion is 
observed in the three versions of the HRG model, compared to the traditional Taylor expansion.  It was observed in Ref. \cite{Borsanyi:2021sxv} that the $T'$-expansion scheme performs considerably better than  Taylor, when expanding lattice QCD thermodynamics to finite $\mu_B$. In that case, it was noticed that $(\partial / \partial T)_{\mu_B} \sim (\partial^2/\partial \mu_B^2)_T$ near and above the chiral crossover temperature, due to chiral criticality scaling. This is an important observation, since the $T'-$expansion coefficients $\kappa_{2}^{B}$ and $\lambda_{2}^{B}$ contain ratios like $\chi_{4}^{B}/{\chi'_{2}}^{B}$, which are almost constant in the vicinity of the phase transition. On the contrary, the lattice QCD Taylor expansion coefficients show peaks and oscillations around the phase transition. Typically, these peaks and oscillations are reflected in similar (unphysical) behavior in the thermodynamic observables expanded to large $\mu_B/T$. However, in the case of the HRG models discussed here, the Taylor coefficients do not show this oscillatory behavior, leading to generically well-behaved thermodynamics even at large $\mu_B/T$ from truncated Taylor expansion. To study a case where the $T'-$expansion schemes have a real advantage compared to Taylor, in the next session we discuss the Cluster Expansion Model.

\subsection{Cluster Expansion Model}
As described in Section \ref{CE}, the first two coefficients in the Cluster Expansion Model are tuned to reproduce lattice QCD results, while all others can be obtained from Eq. (\ref{eq:CEMscale}).
Taylor expansion coefficients can be obtained analytically, and they are in quantitative agreement with lattice QCD results \cite{Vovchenko:2017gkg}.  Therefore, they show the same wiggly behavior as the lattice QCD coefficients, and exhibit a similar scaling between derivatives of the pressure with respect to $T$ and $\mu_B$.

\begin{figure}[!h]
    \centering
\includegraphics[scale=0.28]{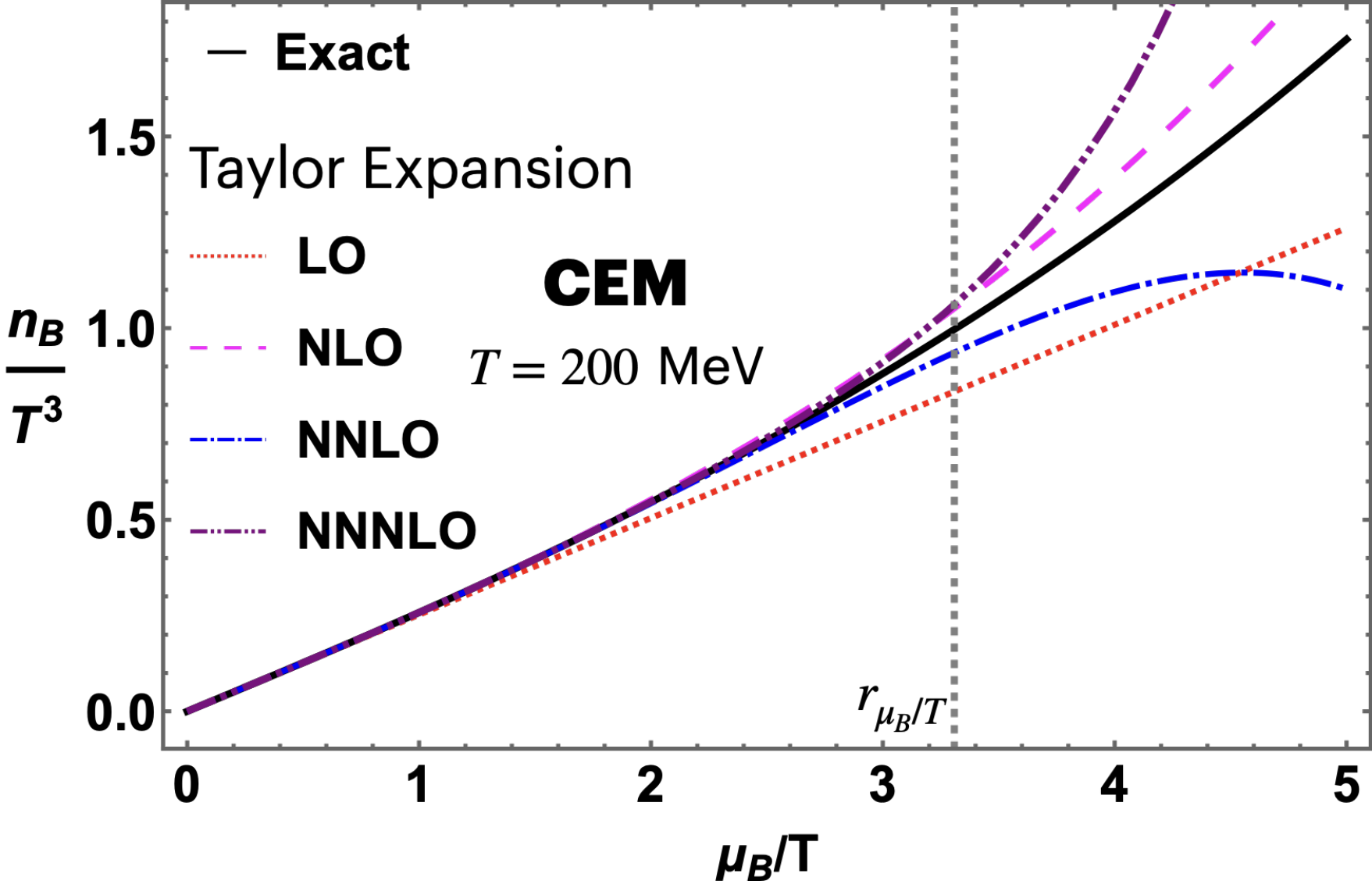}\\
    \centering
\includegraphics[scale=0.28]{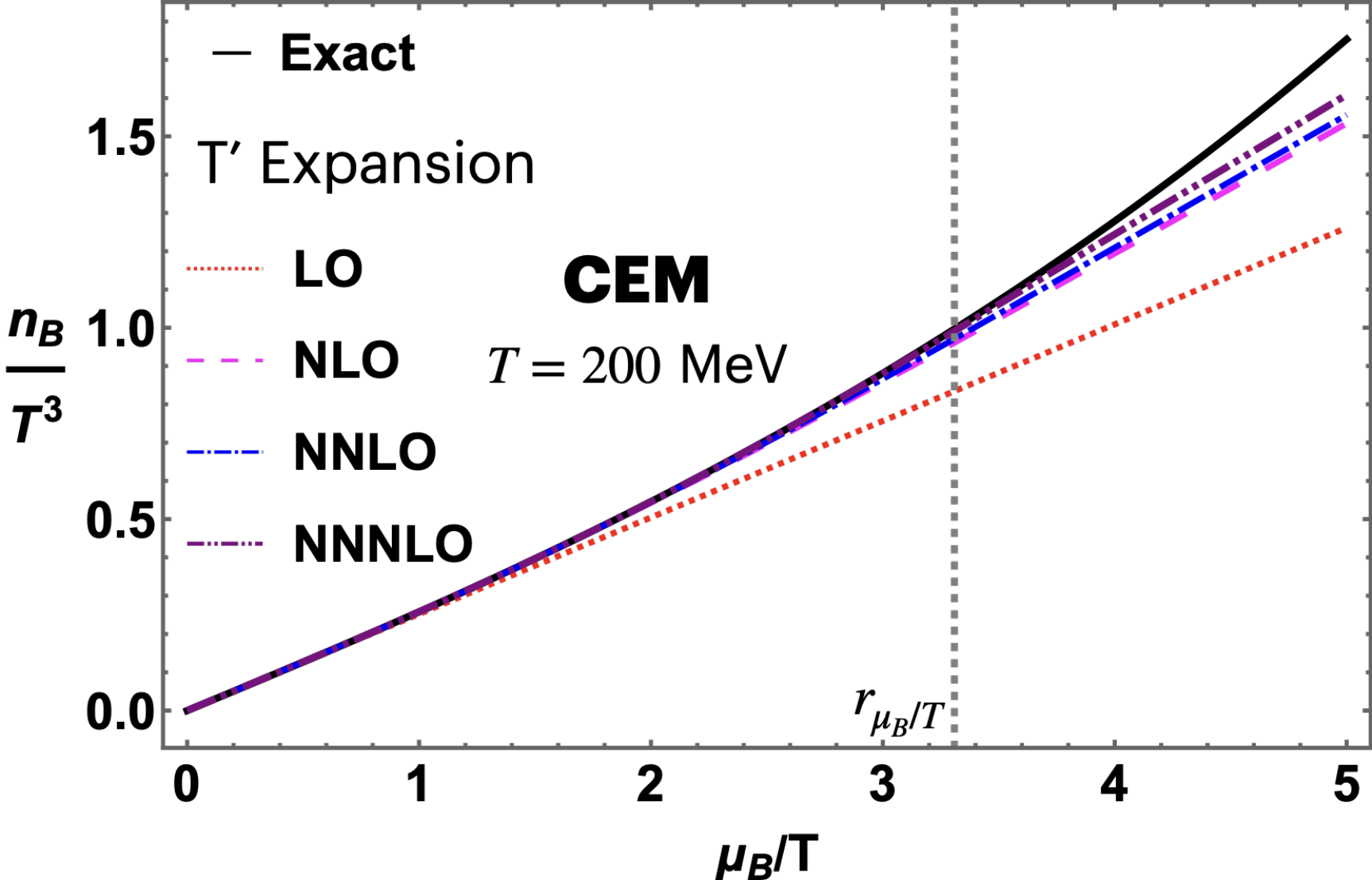}\\
\includegraphics[scale=0.28]{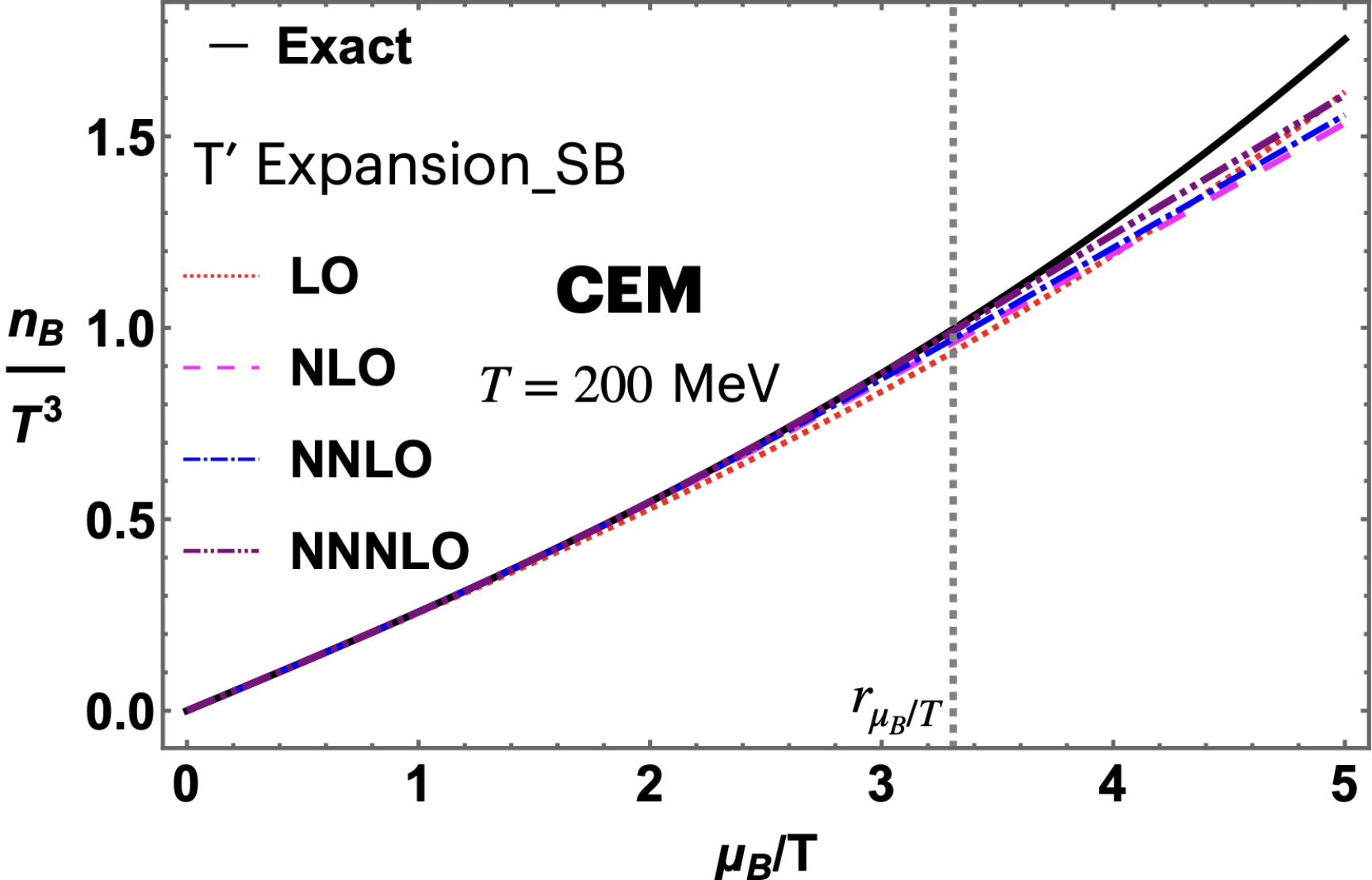} 
\caption{Scaled baryon density as a function of $\mu_B/T$ in the Cluster Expansion Model at $T=200~ \text{MeV}$, where the Roberge Weiss radius of convergence of the Taylor expansion corresponds to $\mu_B/T\approx\pi$, indicated by the black vertical dashed line. The top panel corresponds to the Taylor expansion, the middle panel corresponds to the $T'-$expansion scheme and the bottom panel to the $T'-$expansion with Stefan-Boltzmann limit. In all panels, the solid black line indicates the exact result, while the different dashed color lines correspond to different expansion orders.}
    \label{fig:TExS_200}
\end{figure}

Figure \ref{fig:TExS_200} shows the Taylor expansion (top panel), $T'-$expansion (central panel) and $T'-$expansion with SB limit (bottom panel) applied to the Cluster Expansion Model. As before, the scaled baryon density is plotted as a function of $\mu_B/T$, for $T=200$ MeV. The different curves correspond to the full result (black, solid) and different truncation orders in $\mu_B/T$. It is evident from these figures that, in the case of the CEM, the two versions of the $T'-$expansion scheme allow for a better agreement between the truncated and full results at large chemical potentials $\mu_B/T>3.5$. 
The CEM contains a Roberge-Weiss-like singularity in the complex $\mu_B/T$ plane, which at $T = 200$~MeV yields a radius of convergence for the Taylor expansion $r_{\mu_B/T} \approx \pi$, which can be obtained from  Eq.~\eqref{eq:rmuTCEM}. This value of $\mu_B/T$ is indicated as a vertical dotted line in the panels of Fig. \ref{fig:TExS_200}.

\begin{figure*}[htb]
    \centering
    \includegraphics[width=\linewidth]{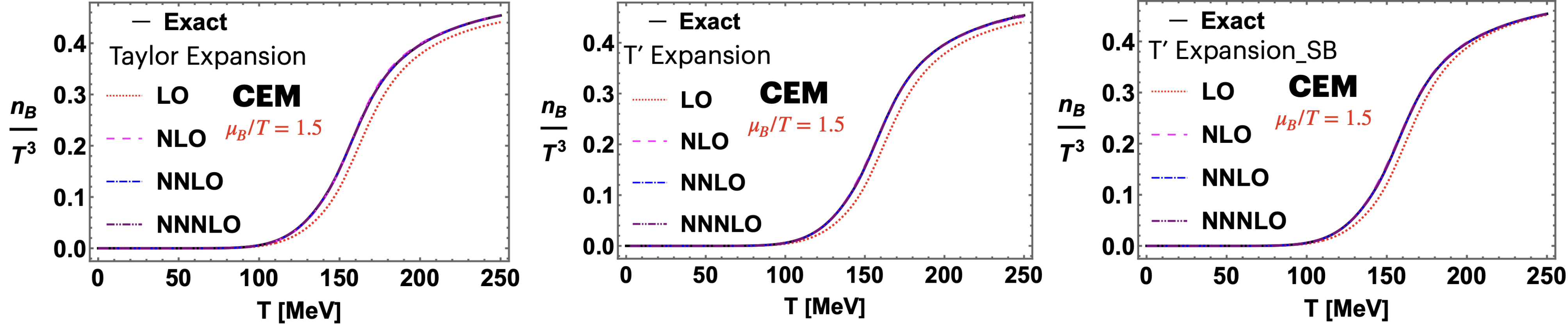}
    \includegraphics[width=\linewidth]{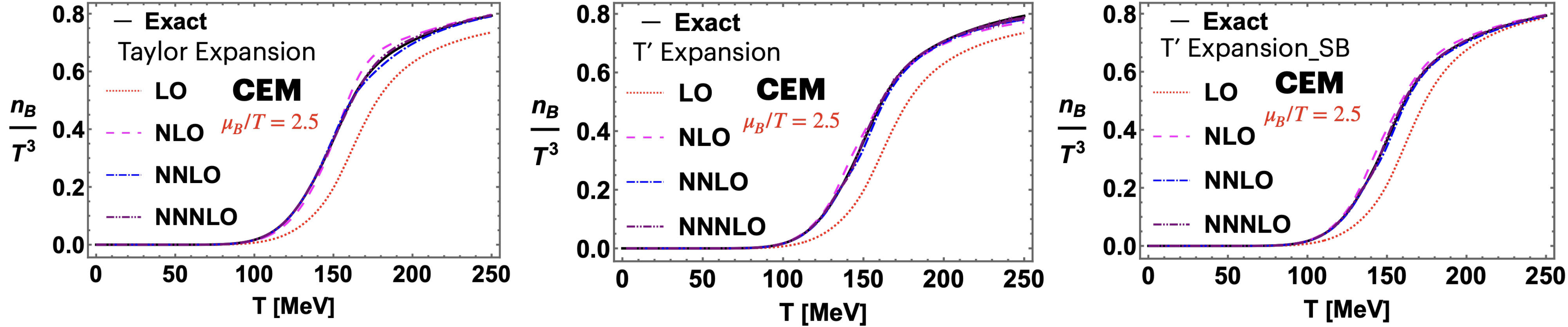}
     \includegraphics[width=\linewidth]{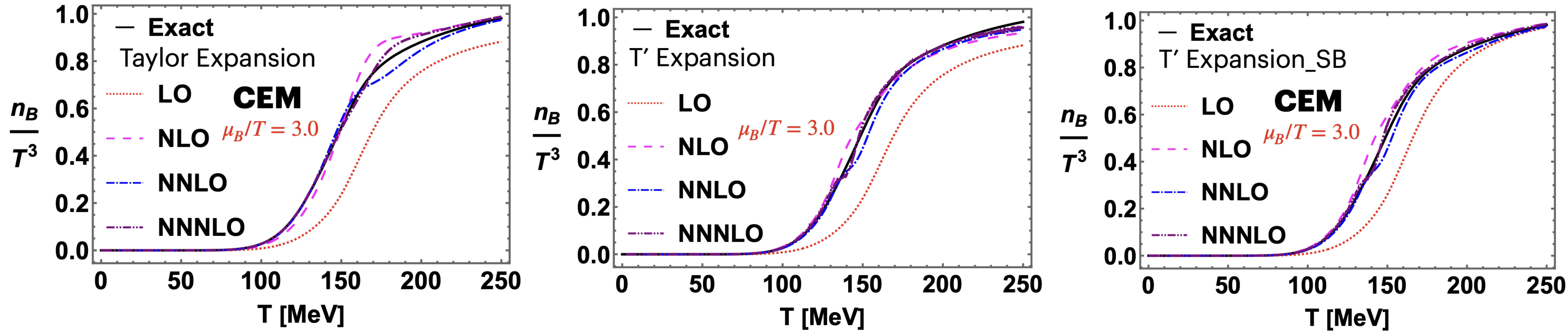}
    \includegraphics[width=\linewidth]{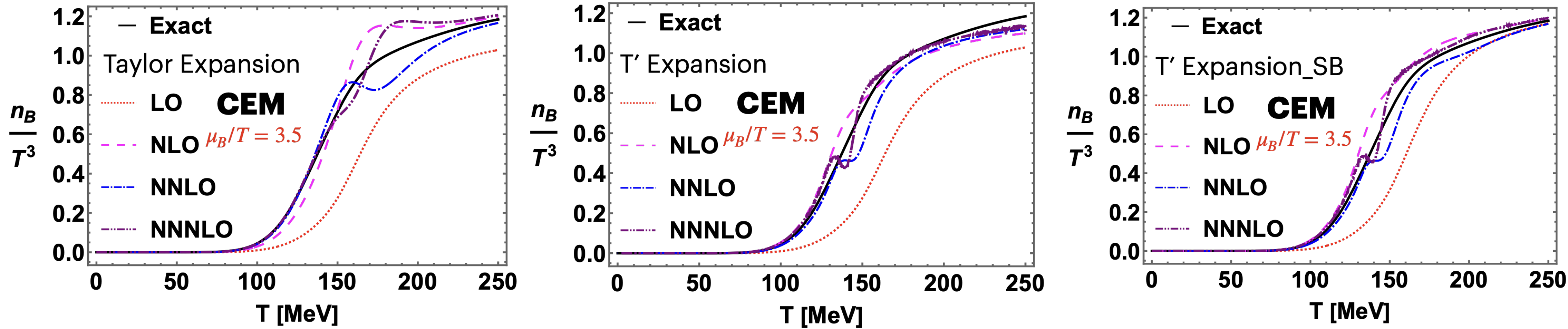}
    \caption{ Scaled net-baryon density as a function of the temperature, for different values of $\mu_B/T$, calculated in the Cluster Expansion Model. The top panels correspond to $\mu_B/T=1.5$, the second row to $\mu_B/T=2.5$, the third row to $\mu_B/T=3$ and the last row to $\mu_B/T=3.5$.
    The left column corresponds to the Taylor expansion, the central column to the $T'-$ expansion scheme and the right column to the $T'-$expansion scheme with SB limit. In all panels, the solid black line shows the full result, while the dashed/dotted lines in different colors correspond to different truncations in powers of $\mu_B/T$.
    }
    \label{Taylor_TExS_SB}
\end{figure*}

To better compare the performance of the Taylor expansion and the two versions of the $T'-$expansion scheme, in Fig. \ref{Taylor_TExS_SB} we show the scaled baryon density in the CEM as a function of the temperature, for $\mu_B/T=1.5$ (top panels), $\mu_B/T=2.5$ (second panels from the top), $\mu_B/T=3$ (third panels from the top), $\mu_B/T=3.5$ (bottom panels).
The left panels show the Taylor expansion results, the central panels the $T'-$expansion scheme and the right panels the $T'-$expansion scheme with SB limit. In all cases, the full result is indicated as a solid black line, while different colored lines show different orders in the expansion truncation in $\mu_B/T$.
It is clear from this Figure that, in the case of the CEM, the $T'-$expansion schemes perform better than Taylor at high $\mu_B/T$, where the Taylor expansion shows large oscillations reflecting the oscillatory behavior of the Taylor coefficients.

The radius of convergence in the $T'-$expansion scheme is not well-defined. Its convergence is expected to coincide with that of the Taylor expansion in Eqs.~\eqref{Eq:Tprime} and~\eqref{eq:TprimeSB} for $T'$ as a function of $\mu_B/T$, given that the $T'-$expansion scheme does not contain other possible singularities (it only defines a relationship between $\chi_1^B(T,\hat{\mu}_B)/{\hat{\mu}_B}$ and $\chi_2^B(T,\hat{\mu}_B = 0)$, which is a regular smooth function at $\hat{\mu}_B = 0$).
However, due to the absence of an explicit form for the $T'(\mu_B/T)$ dependence, it is challenging to determine the value of the convergence radius in $T'-$expansion schemes.
We have performed estimations of the radius of convergence using the leading four expansion terms and the Mercer-Roberts estimator~\cite{mercer1990centre}~(and its modified version~\cite{Giordano:2019slo}) in the HRG models.
These leading-order estimates tend to place the radius of convergence in both $T'-$expansion schemes slightly above the same-order estimates of the standard Taylor expansion, which is consistent with the general observations discussed above regarding the performance of the different schemes.

\section{\label{Summary}{Summary}}

The $T'-$expansion scheme for lattice QCD thermodynamics was constructed based on the observation that 
the leading-order expansion coefficient $\kappa_2^B$~(or $\lambda_2^B$ in the $T'-$expansion scheme with corrected SB limit), 
is approximately constant or linear in $T$ in the region where the transition line is expected, respectively. 
This means that $ \chi_2^{B'} \sim \chi_4^B $ and  $\chi_2^{B''} \sim \chi_6^B $.  
This happens because $\frac{\partial }{ \partial T} \sim \frac{\partial^2}{ \partial \mu_B^2}$, a scaling which is true in the chiral limit, and 
 approximately preserved at physical quark masses
as seen from lattice QCD results. 
Besides, the next expansion coefficients $\kappa_4^B$ and $\lambda_4^B$ are consistent with zero, which makes their contribution at moderate chemical potential almost negligible.
Effective theories that exhibit the same scaling properties are expected to work well under the $T'-$expansion scheme.

In this work, we investigated the convergence properties of the 
$T'$-expansion scheme, with and without the Stefan-Boltzmann limit correction, compared to the Taylor expansion. The expansion behavior of the scaled baryon density is analyzed in the low-temperature regime using the Hadron Resonance Gas model, and at high temperature using the Cluster Expansion Model and the Holographic model, whose results are not shown as they are similar to those of the Cluster Expansion Model.
It is observed that the $T'$-expansion scheme shows better convergence behavior than Taylor when applied to the Cluster Expansion Model since this model largely preserves the scaling $\partial/\partial T \sim \partial^2/\partial\mu_B^2$. This is not true for the HRG model: $\kappa_2^B$ and $\lambda_2^B$ are monotonically increasing functions in the Ideal case and non-monotonic functions of the temperature for EV- and vdW-HRG models. In all these cases, the $T'-$expansion scheme performance is comparable with the one of the Taylor expansion: a similar number of expansion terms is needed, to cover a similar range in $\mu_B/T$ with a reasonable agreement with the full result.
We also observe that while $T^{\prime}$ expansion offers similar or superior performance to Taylor expansion at low orders, its advantages at higher orders appear to be limited.
Therefore, there is no evidence from model studies that evaluating high-order $T'$-expansion coefficients, which is very expensive to do with lattice QCD, will offer substantial advantages in describing the equation of state at finite baryon density.

\begin{acknowledgments}
This material is based upon work supported by the National Science Foundation under grants No. PHY-2208724, PHY-1654219 and PHY-2116686, and within the framework of the MUSES collaboration, under grant number No. OAC-2103680. This material is also based upon work supported by the U.S. Department of Energy, Office of Science, Office of Nuclear Physics, under Award Number
DE-SC0022023 and by the National Aeronautics and Space Agency (NASA) under Award Number 80NSSC24K0767.

\end{acknowledgments}
\bibliography{apssamp}

\end{document}